\documentclass[mathleft,fleqn,%
]{an}
\def\ferre{{\tt FER\reflectbox{R}E}}
\usepackage{graphicx}
\usepackage[varg]{txfonts}
\begin{document}

\Pagespan{1}{}
\Yearpublication{2015}%
\Yearsubmission{2015}%
\Month{0}%
\Volume{999}%
\Issue{0}%
\DOI{asna.201400000}%

\title{Automated Pipelines for Spectroscopic Analysis}
\author{C. Allende Prieto\inst{1,2}\fnmsep\thanks{Corresponding author:
        {callende@iac.es}}
}
\titlerunning{Automated Pipelines for Spectroscopic Analysis}
\authorrunning{C.}
\institute{
Instituto de Astrof\'{\i}sica de Canarias, C/ V\'{\i}a L\'actea s/n, 38205 La Laguna, Tenerife, Spain
\and 
Universidad de La Laguna, Departamento de Astrof\'{\i}sica, 38206 La Laguna, Tenerife, Spain
}

\received{XXXX}
\accepted{XXXX}
\publonline{XXXX}


\abstract{
The Gaia mission will have a profound impact on our understanding
of the structure and dynamics of the Milky Way. Gaia is providing an exhaustive census 
of stellar parallaxes, proper motions, positions, colors and radial velocities,
but also leaves some flaring holes in an otherwise complete data set.
The radial velocities measured with the on-board high-resolution spectrograph 
will only reach some 10\% of the full sample of stars with astrometry and 
photometry from the mission, and detailed chemical information will be 
obtained for less than 1\%. Teams all over the world are organizing 
large-scale projects to provide complementary radial velocities and
chemistry, since this can now be done very efficiently from the ground 
thanks to large and mid-size telescopes with a wide field-of-view and 
multi-object spectrographs. As a result, automated data processing
is taking an ever increasing relevance, and the concept is applying
to many more areas, from targeting to  analysis. In this paper, 
I provide a quick overview of recent, ongoing, and upcoming spectroscopic 
surveys,  and the strategies adopted in their automated analysis pipelines.
}

\maketitle

\section{Introduction}

Gaia was launched on December 19, 2013. Science operations for the mission started 
on July 2014, and its suite of instruments will hopefully continue to gather data 
continuously during the nominal mission lifetime of five years. Gaia expands 
the global all-sky measurement of stellar positions made by the 
Hipparcos mission, which flew between 1989 and 1993, from 0.1 kpc to 20 kpc, 
increasing the number of targets by 4 orders of magnitude with a precision improved 
by 2 orders of magnitude. In addition to astrometry (positions, proper motions, 
and trigonometric parallaxes),
Gaia includes a pair of spectrophotometers, BP-RP, covering the range 360-1000 nm, 
and a high-resolution spectrograph, the RVS, observing in the range 847-874 nm. These
instruments provide spectral energy distributions for all the stars with astrometry
(about $10^9$ stars down to $V \sim 20$) and radial velocities for the brightest 10\% of them.

Without a doubt, Gaia's data will revolutionize our understanding of 
the structure, formation and evolution of the Milky Way, galaxies 
in general, and various
aspects of stellar formation and evolution. But, if extremely rich,
the Gaia data set is quite limited regarding chemistry information, since spectral lines
are not resolved in the BP-RP observations, and the 
signal-to-noise
ratio of the RVS spectra is too low to measure chemical abundances for stars
fainter than about $V=12$. Furthermore, the radial velocity information will
be limited to stars brighter than $V\simeq16$.

This situation has triggered the reaction of the astronomical community, 
who is organizing complementary projects to perform spectroscopy from 
the ground. Three massive high-resolution ($R\equiv \lambda/\delta\lambda 
\sim 20,000$) projects are currently underway:  APOGEE, Gaia-ESO, and GALAH. 
There are also
ongoing efforts at lower resolution ($R\sim 2000$), the SDSS and LAMOST, 
as well as the RAVE survey, that uses intermediate resolution 
but a more limited (RVS-like) spectral coverage. The main targets and 
instrumental characteristics of these projects and many others planned 
for the near future are summarized in \S 2. 

The massive data sets being produced call for
a scale of automation never seen  before, from target selection to
instrumental configuration, to data acquisition, data reduction, and 
even  analysis. Data products from these surveys
are far more advanced than calibrated spectra, and involve the use
of physical models, crossing the line between  
actual observations and their theoretical interpretation. Section 3
will make a quick overview of the most usual methodologies involved in 
a basic analysis of stellar spectra, and \S 4 will
glance at the construction of models. Section 5 mentions some of the 
most popular algorithms and codes adopted for the automated derivation of
atmospheric parameters and chemical abundances.

The architecture of the analysis pipelines for the different surveys
can vary wildly, and Section 6 will describe the ones adopted
by three of the most relevant projects.

\section{Ongoing and future ground-based spectroscopic surveys}

\subsection{Current projects}     

At low resolution, the largest projects by far are 
the Sloan Digital Sky Survey (SDSS) and the
Large Area Multi Object fiber Spectroscopic Telescope (LAMOST).

The SDSS has been running for about 15 years, using a dedicated
2.5-m telescope (Gunn et al. 2006) at Apache Point Observatory, in
New Mexico. The project established its own 5-band $ugriz$ photometric system, 
and mapped a large fraction of the Northern sky. Targets for 
spectroscopy are selected, mainly from photometry, to fulfill a variety of
science objectives. The original SDSS project (1998--2005; 
York et al. 2000), 
the Baryonic Oscillations Spectroscopic Survey (BOSS 2009-2013; 
Dawson et al. 2013, Eisenstein et al. 2011), and 
its sequel eBOSS (2014-2020; Zhao et al. 2015), target galaxies and quasars, but
included some stars for addressing 
particular research topics, and F-type halo sub-dwarfs 
for calibration. The SEGUE and SEGUE-2 projects (2005-2008; see Yanny et al.
2009)
focused  on stars. Altogether, the SDSS archive has low-resolution spectra
for about a million stars in the range $14<V<21$, all publicly
accessible through their regular data releases (the latest was DR12;
Alam et al. 2015),  and continues operating. 

The two original double-arm SDSS spectrographs  
used in the original survey, SEGUE, 
and SEGUE-2, were upgraded in 2009 to enhance throughput, resolution, 
and spectral coverage (Smee et al. 2013). Both the original and
upgraded instruments share a resolving power about 2,000 and
broad spectral coverage in the optical (380-960 nm for the original
and 360-1000 nm for the upgraded version). These instruments are fiber 
fed from plug-plates mounted on the 3-degree focal plane of the telescope 
and accommodate 640 3-arcsecond (original instrument) or 1000 
2-arcsecond diameter (upgraded) fibers simultaneously.  Since
2011 the SDSS incorporates the APOGEE high-resolution H-band 
spectrograph, which 
is described below.

Inspired by the SDSS, the Large Sky Area Multi-Object Fibre Spectroscopic Telescope
(LAMOST) started regular operations in
2011. This telescope, which has an original design and
an effective aperture in the range 3.6-5.9m, is used together with 
 an advanced robotic fiber positioner
to acquire up to 4000  objects per exposure. The fibers feed 16 spectrographs, 
typically set up to provide a resolving power about 1500 and 
broad spectral coverage between 370 and 900 nm. A recent dedicated issue of
the journal  Research in Astronomy and Astrophysics has described the results
of the Milky Way observations from LAMOST. The first public data release
took place earlier in 2015 (Luo et al. 2015) and included nearly 3 million 
spectra of Northern stars, most in the range $13<V<19$.

There are currently three massive ongoing projects providing stellar
spectra over a large area of the sky with a resolving power of $\sim 20,000$, or
about ten times higher than SDSS/SEGUE or LAMOST: APOGEE
(Majewski et al. 2015; also part of SDSS), Gaia-ESO (Gilmore et al. 2012),
and GALAH (Zucker et al. 2013). 

The Apache Point Observatory Galactic Evolution Experiment
(APOGEE) started gathering data in 2011 and it couples a 
300-fiber H-band (1.5-1.7 $\mu$m) spectrograph to the SDSS 2.5m telescope. After
three years of operations, the project made a full data release in January 2015
including spectra for more than 150,000 Northern stars in the range
$8<H<14$ (or $10<V<17$), most of them
red giants  in the Galactic disk, but also reaching into the 
Galactic bulge and the halo (Holtzman et al. 2015). APOGEE observations will
continue at least until 2020, and a replica of the APOGEE spectrograph is being
built and will perform parallel observations from Las Campanas starting 
in Fall 2016. The H-band is rich in atomic and molecular
lines, and the APOGEE spectra recover atmospheric parameters and
abundances of 15 elements for red giant stars.

\begin{table*}
\centering
\caption{Performance of ongoing survey instruments.}
\label{tlab}
\begin{tabular}{ccccccc}\hline
Project/Instrument & R & $\Delta\lambda/<\lambda>$ & S/N &  P &  N &  V \\ 
                   &   &                           &     &    &    & mag \\
\hline
Gaia BP-RP& 100    &  1    & 30  & $3,000$  & 10$^9$       & 8-20  \\
LAMOST   & 1,500  &  1    & 20  & $30,000$  & 10$^7$         & 12-18 \\
SDSS     & 2,000  &  1    & 30  & $60,000$  & 10$^6$         & 14-20  \\
RAVE     & 7,500  & 0.05  & 50  & $18,750$  & $500,000$  & 8-14  \\
Gaia RVS & 11,500 & 0.05  &  2  & $1,150$          & 10$^8$         & 5-17 \\
Gaia-ESO & 20,000 & 0.12   & 80  & $192,000$   & $100,000$         & 14-19 \\
APOGEE   & 22,500 & 0.12  & 100 & $270,000$   & $400,000$  & 10-17 \\
GALAH    & 28,000 & 0.15  & 100 & $420,000$        & $10^6$      & 10-17 \\
\hline
\end{tabular}
\end{table*}

The Gaia-ESO survey uses general-purpose ESO facilities, namely
one of the VLT 8m telescopes (Kueyen) and the FLAMES instrument,
which feeds medium and high-resolution spectrographs.
The project started at the end of 2012
and will extend at least to 2017. Most stars are observed at
$R\sim 20,000$ with GIRAFFE, but about 10\% are brighter stars
fed to UVES, with $R\simeq 50,000$. As with other ESO Public Surveys,
the raw data become immediately available, and reduced spectra, 
atmospheric parameters, and chemical abundances are released at a slower pace.

The GALAH survey uses the 4m Anglo-Australian telescope and
a custom-made 4-arm spectrograph (HERMES), with the resolving power
set to $R\sim 28,000$ to target 400 objects
per exposure, and measure abundances for up to 30 elements. 
The project pursues collecting spectra for a million stars in the
range $12<V<14$.
The instrument was commissioned at the end of 2013 and
has already obtained spectra for more than 100,000 stars. Its
first public data release is planned for mid-2016. 

With an intermediate resolving power $R\simeq 7000$ and limited
spectral coverage (841-880 nm), the Radial Velocity Experiment (RAVE) 
obtained data between 2003 and 2013 using the 1.2m UK Schmidt Telescope at the 
Australian Astronomical Observatory for about 500,000 Southern stars with 
$9<V<14$ (Steinmetz et al. 2006). 
Their latest data release took place in 2013 (Kordopatis et al. 2013) 
and included parameters, distance estimates, and abundances for up to six elements 
for a large fraction of the targets, but no spectra. The next data release
is planned for 2016.

Table 1 summarizes the main parameters for each of the projects described
above. Fig. 1 illustrates the situation. In addition to  
the number of observed stars N, their approximate $V$ magnitude range,
and the resolving power of the instruments $R$, 
I have computed an additional quantity, the {\it power to resolve}, $P$, that 
combines the resolving power with other important factors, namely the 
relative spectral coverage $\Delta\lambda/<\lambda>$, and the signal-to-noise ratio,
 in an attempt to measure of the information content per observation
\begin{equation}
P = R  \left(\frac{\Delta\lambda}{<\lambda>}\right) \left(\frac{S}{N}\right).
\end{equation}
\noindent Since $P$ is proportional to $R$, there is correlation between
the two which the RVS does not share, mainly due to its atypically low signal-to-noise
ratio and spectral coverage.  As  one may expect, $P$ is anticorrelated with the
number of targets $N$: the more information per spectrum,
the harder it gets to observe a large number of objects.

\begin{figure*}
\includegraphics[height=160mm,angle=90]{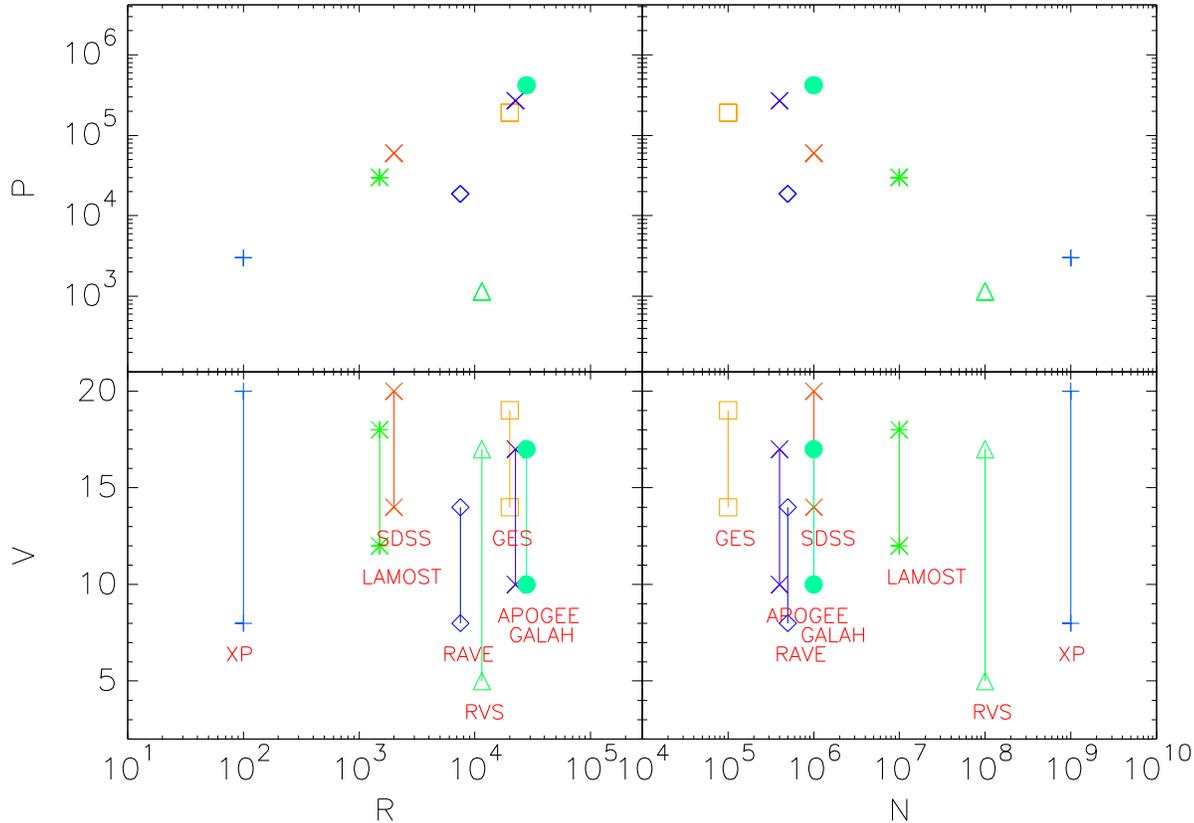}
\caption{Recent and ongoing spectroscopic surveys 
as a function resolving power $R$, power-to-resolve $P$ (see text), 
number of targets $N$, and $V$ magnitude range.}
\label{f1}
\end{figure*}

\subsection{The future}

The existence of multiple projects carrying out spectroscopic surveys
of the sky, and stars in the Milky Way in particular, has not precluded
additional projects to get organized.

WEAVE for the 4m WHT in La Palma (Dalton et al. 2014) 
and  4MOST for the 4m VISTA telescope at Paranal  (de Jong et al. 2014) 
will provide  multi-object medium ($R\sim 5000$) and high-resolution 
($R\sim 20,000$) optical spectroscopy from the 
Northern and Southern hemispheres, respectively, and embark in massive 
surveys including Milky
Way stars. WEAVE has a robotic fiber positioner handling 1000 
fibers and a single spectrograph that can work in medium or high-resolution modes,
while 4MOST will have about 2400 fibers, part feeding a medium-resolution spectrograph and
part feeding a high-resolution one. With similar medium or high resolution modes, 
on a larger telescope, 
the ESO 8m VLT, MOONS will provide near-IR coverage
(0.7--1.7 $\mu$m) for fainter targets (Cirasuolo et al. 2014).
These three instruments are planned to start operations in 2018-2021.

Another project to keep an eye on is the Hobby-Eberly Telescope Dark Energy
Experiment (HETDEX; Hill et al. 2008), 
which uses an innovative massive instrument, VIRUS. This instrument 
feeds light from 30,000 fibers statically arranged into 75 Integral Field
Units to 150 spectrographs. This experiment is designed mainly 
for cosmology, but about 200,000 stars down to $V\simeq 22$ will be observed 
along with galaxies in a 60-square degree region in the vicinity of the Big Dipper.
The low resolving power ($R\sim 700$) is compensated with a spectral range
that reaches into the near-UV (350-550 nm), where a higher density of stellar 
absorption lines helps to measure radial velocities or estimating 
stellar metallicities. HETDEX will start in 2016.

Finally, the Dark Energy Spectroscopic Instrument  (DESI), will involve a robotic
positioner with 5000 fibers and 10 three-arm spectrographs working at 
medium resolution ($2000< R <5500$, depending on wavelength)  
over the range 360-980 nm (Levi et al. 2013). The project employs the 4m Mayall
telescope at Kitt Peak, aims to start operations in 2019-2020,
and it is a good candidate to populate the exciting 
upper-right area in the upper-right panel of Figure 1.

\section{Analysis methodology}

The most basic analysis is spectral classification. The classical MK system
is still in use, but many alternative machine-based schemes have been proposed
(see Bailer-Jones 2002 for a review). These involve cluster analysis techniques 
such as {\it K-means}, optimization techniques, or mixture models (see, e.g., Everitt et al. 2011), 
or artificial neural  networks (Bailer-Jones et al. 2002), among others.

More interesting than classification is parameterization. Spectra
depend mainly on the fundamental atmospheric parameters, the stellar effective
temperature ($T_{\rm eff}$), surface gravitational acceleration ($\log g$),
and its chemical composition (simplified as a single {\it metallicity}, [Fe/H], 
the logarithm of a scale factor that applies to the solar mixture). In this paper
we will refer to this set of parameters with the letter {\bf p}.
Parameterization is either performed based on model spectra, or observed ones.
In any case, when observed templates are used, somebody has to assign parameters
to them based again on model spectra.

\section{Model spectra}

Model spectra can be computed under a given set of approximations. Traditionally,
hydrostatic equilibrium, energy conservation, and local thermodynamical equilibrium
are assumed in order to calculate a model atmosphere. 
Then detailed spectra are computed by solving the radiative transfer equation
with detailed opacities for lines and continua. 

Large sets of classical model
atmospheres are available, and  sparse grids of more detailed hydrodynamical models 
are becoming available (Ludwig et al. 2009, Trampedach et al. 2013).
An overview of the available sources of model atmospheres, opacities, 
and radiative transfer codes is given in Allende Prieto (2016).

\section{Algorithms and codes}

Once we are ready to compute model spectra, we can focus on the task
of identifying the algorithm to find the set of model parameters {\bf p}
that best reproduces any given observed stellar spectrum. The 
possibilities are endless and only some of the most commonly employed
techniques will be mentioned here.

The traditional methods focus on quantifying the strength of absorption
lines by measuring their {\it equivalent widths}. Many lines are usually 
available for transitions of two iron ions, e.g. atomic and singly-ionized iron
in the case of late-type stars, and forcing the inferred iron abundance to
be independent of the line excitation energy, the line strength, and the
ion can constrain the atmospheric parameters. This technique, however, is
limited to fairly high spectral resolution data. 

Projection algorithms take input spectra and identify functional relationships
that map those onto the desired parameters. Neural networks fall in this category,
and so does the MATISSE algorithm by Recio-Blanco et al. (2006), or "the Cannon"
(Ness et al. 2015).

In most cases there is a unique solution and local optimization techniques 
such as the Nelder-Mead algorithm (Nelder \& Mead 1965),  the
Levenberg-Marquardt algorithm (Marquardt 1966), or the conjugate gradient
method (see, e.g., Shewchuk 1994) can be very efficient. 

When the multidimensional {\bf p} space shows a complex landscape with
multiple local minima we may need to put up an extra effort using 
global optimization algorithms, such as annealing (Kirkpatrick 1984)
 or genetic algorithms (Goldberg 1989).
Bayesian techniques, coupled or not to Markov-Chain Monte Carlo chains 
to optimize the number of function evaluations, can also be used to search 
for the optimal solution, with the advantage of having the possibility of 
folding-in external information about the sample we are observing (see, e.g., Lee 2013).

\section{Pipeline architecture}

A spectroscopic analysis pipeline is a software package that
takes fully reduced spectra as input, and derives physical information
such as radial velocities, atmospheric parameters, or chemical abundances,
from them. Most outputs depend on models, i.e. are model-dependent. 
The connection between models and parameters can be implicit, through
relationships that have been determined beforehand, or explicit,
through the direct calculation of synthetic spectra during the pipeline
execution. As mentioned above, one may use libraries of existing 
observations, but the parameters assigned to those will be ultimately
tied to model atmospheres and synthetic spectra.

The architecture of a pipeline depends on the quantity and quality of
the input data, which sets what can be extracted and how much information
the pipeline needs to digest and at which speed. It will depend on whether
the spectra themselves, or derived quantities, such as equivalent widths
or spectral indices, are used in the evaluation of the merit function that
defines what are the most likely values for the sought-after parameters.

A pipeline may seek multiple parameters at once, or sequentially.
It may adopt a single optimization algorithm or a number of them.
It may also embrace a single model set (model atmospheres, opacities, etc.)
or several of them.  Pipelines can be developed specifically for 
a given instrument, survey, or project.
But in some instances they can be very general and be used in multiple ones.

It is not always obvious which choices are best and whether there is
a recipe that can be applied in most situations. In this paper, I will
discuss some of the choices adopted for three particular surveys in which
I have been involved: APOGEE, Gaia-ESO, and SDSS-SEGUE.

\subsection{Example 1: APOGEE}

The APOGEE pipeline (ASPCAP; Garc\'{\i}a P\'erez et al. 2015) 
uses the chi-squared between observed and model spectra
to decide on what are the most likely values for the parameters it searches:
radial velocities, atmospheric parameters and chemical abundances. With 
a resolving power of about 20,000 and a spectral coverage between 1.5 and 1.7 $\mu$m,
computing the chi-squared implies a loop over $10^4$ wavelengths. The
APOGEE data are very homogeneous -- all spectra are acquired with the 
same instrument/setting.

The pipeline determines 6 or 7 parameters simultaneously for each APOGEE spectrum:
$T_{\rm eff}$, $\log g$, micro-turbulence, [M/H], [C/M], [N/M], and [$\alpha$/M].
The remainder of the chemical abundances are derived in a second step, one element
at a time, holding constant the parameters derived in the previous stage. The rationale 
for pursuing 
the carbon, nitrogen, and $\alpha$-element abundances in the first optimization 
is that these elements
can have a critical effect on the derivation of the main atmospheric parameters
($T_{\rm eff}$, $\log g$ and metallicity [M/H]), through their effect on the
equation of state or the opacity, mainly through molecular absorption (CN, OH or CO)
or contributing free electrons.

The APOGEE pipeline has only one algorithm for deriving atmospheric
parameters and abundances, currently the Nelder-Mead algorithm, as implemented
in the code \ferre, written in FORTRAN90.
\ferre\ is open source\footnote{Available from http://hebe.as.utexas.edu/ferre}
and adopts a strategy to analyze spectra that is applicable to virtually
any type of spectroscopic data. The code is optimized to run on large samples,
and evaluates model spectra by interpolating in a grid of pre-computed model fluxes,
which can be compressed using Principal Component Analysis. In the APOGEE
pipeline FERRE is wrapped within a complex book-keeping software written
in IDL, which prepares the spectra, launches FERRE jobs,
and sorts and packs the output. 

The effective temperatures and abundances obtained are not far from 
those expected, while gravities are more affected by systematic errors. 
Offsets between reference data for open and globular clusters
and stars with their properties derived from oscillations are tracked,
modeled with simple functions, and calibrated out.  

Following the model generally used in previous SDSS pipelines, the APOGEE
pipeline software is version-controlled in a project server, and the
pipeline itself runs on project computers in an automated fashion. 
This allows the analysis of new data to be done consistently, and makes it possible
to reproduce the results. The software used in the analysis becomes publicly
available from the SDSS servers together with the data they have been
run on at each public data release of the SDSS (see, e.g. Alam et al. 2015).

\subsection{Example 2: Gaia-ESO}

The Gaia-ESO analysis pipeline is actually a suite of pipelines run
by individual teams at different locations (Smiljanic et al. 2014; Recio-Blanco
et al. 2014). Each team ({\it node})
develops its software and is in charge of maintaining it. A set of
common guidelines regarding the basic input data used to model
 spectra (model atmospheres, atomic and molecular data, reference
solar abundances, etc.) are given, but the actual choices
of algorithms, codes, and strategies to extract the information from
the spectra are left to the nodes.

The Gaia-ESO Public Survey employs both the GIRAFFE and UVES
spectrographs, with roughly 90\% of the data coming from the former.
There are a few nodes involved in the analysis of the GIRAFFE data, which 
are coordinated as a working group,
and the parameters from them are mapped onto a common scale and then
averaged. In the case of UVES data, the corresponding working group includes 
over a dozen nodes, and the results from them are averaged out using weights 
defined upon their measured performance on a set of benchmark stars 
(Heiter et al. 2015; Jofr\'e et al. 2015).

In addition to the working groups dealing with GIRAFFE or UVES data
for late-type {\it normal} stars, there are additional working groups
devoted to the analysis of hot stars, or chemically peculiar stars.
An additional working group enforces a certain degree of homogeneity 
across independent working groups.

The GIRAFFE data for any given star typically include  two spectral
settings, each a few tens of nanometers wide, with a resolving power
about 20,000. Not all stars are observed with the same settings, mainly
to optimize the return for stars in clusters. 
The UVES spectra have a higher resolving power (more than 
twice as high) and broad spectral coverage. In most cases, 
the atmospheric parameters are derived
first, using optimization or projection techniques, and abundances
are determined in a second stage, once the parameters from the different
nodes have been combined into a single set. 

A large fraction of the software used in Gaia-ESO existed before the
survey began, and its performance and behavior was well understood.
However, since the software is not in general open source, and is not kept
under a common software control repository, traceability is  limited 
and operations are not streamlined.

\subsection{Example 3: SDSS-SEGUE}

The SEGUE Stellar Parameters Pipeline (SSPP; Lee et al. 2008 and follow-up
papers) 
is also a suite of software
packages aimed at deriving one or several atmospheric parameters from
SDSS-SEGUE stellar spectra. Some of the packages existed long before
the pipeline was assembled, but others were written specifically for 
SDSS-SEGUE. Some are as simple as polynomial relationships that
relate the equivalent width of a hydrogen line with $T_{\rm eff}$,
while others use algorithms to constrain multiple
parameters simultaneously.

After all the codes have been run on a given data set, a set of
established rules decides which results are adopted and averaged 
depending on the region of the parameter space the solution falls into.

The SDSS-SEGUE  spectra have a resolving power of about 2,000, but have
a broad wavelength coverage. The spectra are fairly uniform, with the
exception of the upgrade of the spectrographs in 2008 for BOSS. 
The focus of the pipeline is to measure
atmospheric parameters ($T_{\rm eff}$, $\log g$\ and [M/H]), although
the overall $\alpha$-element enhancement and the carbon abundance
are also determined. Quality assurance is based on results for clusters, but no
attempt is made to empirically calibrate the outputs from the SSPP.

Similar to the APOGEE case, the SSPP is maintained under version control
on SDSS servers, where it lives and runs, and made publicly available
in sync with public data releases. The
development of the SSPP has slowed down in recent years, and new 
efforts have appeared to analyze the SDSS optical spectra from the
upgraded BOSS spectrographs
(Allende Prieto et al. 2014; Fern\'andez-Alvar et al. 2015).

\section{Summary and conclusions}

The strategies and architectures of existing spectroscopic
analysis pipelines are quite varied. In general, software
pipelines that are open source and applied to multiple
data sets are desirable. Other considerations for designing
a pipeline are 
\begin{itemize}
\item ease of implementation, 
\item computational demands, 
\item performance
(in terms of precision and accuracy, regarding  
 output parameters and their uncertainty),
\item repeatability (software and configuration files must be
	version tracked)
\item clarity and traceability of the results (ease to identify
	what parts of the spectrum are driven by a given parameter).
\end{itemize}

The importance of developing software under version control
that runs and it is maintained at a given location cannot be
overemphasized.  Otherwise repeatability and traceability are compromised.
It is probably a good strategy to focus on one or few algorithms, 
implemented afresh and thoroughly tested, rather than "as many as you
can get", given the limited time available to understand the behavior
of each algorithm (and their average results). Multiple algorithms can 
only provide estimates of
systematic errors if truly independent, i.e. when independent
atomic/molecular data, model atmospheres, synthesis
codes, etc. are used.

\acknowledgements

My research has been supported by the Spanish MINECO (grant
AYA2014-56359-P).
I am grateful to the organizers of the WE Heraeus Seminar
"Reconstructing the Milky Way's History: Spectroscopic surveys, 
Asteroseismology and Chemo-dynamical models"
and to the Wilhelm und Else Heraeus-Stiftung for their kind invitation and
traveling support. An anonymous referee provided useful comments that
improved the presentation. 
  

%

\begin{thebibliography}{}

\bibitem[Alam et al.(2015)]{2015ApJS..219...12A} Alam, S., Albareti, F.~D., 
Allende Prieto, C., et al.\ 2015, \apjs, 219, 12 


\bibitem[]{} Allende Prieto, C., 2016, submitted to Living Reviews
in Solar Physics

\bibitem[Allende Prieto et 
al.(2014)]{2014A&A...568A...7A} Allende Prieto, C., Fern{\'a}ndez-Alvar, E., Schlesinger, K.~J., et al.\ 2014, \aap, 568, A7 

  
\bibitem[Bailer-Jones(2002)]{2002adaa.conf...83B} Bailer-Jones, C.~A.~L.\ 
2002, Automated Data Analysis in Astronomy, 83 

\bibitem[Bailer-Jones et al.(2002)]{2002adaa.conf...51B} Bailer-Jones, 
C.~A.~L., Gupta, R., 
\& Singh, H.~P.\ 2002, Automated Data Analysis in Astronomy, 51 

\bibitem[Cirasuolo et al.(2014)]{2014SPIE.9147E..0NC} Cirasuolo, M., 
Afonso, J., Carollo, M., et al.\ 2014, \procspie, 9147, 91470N 



\bibitem[Dalton et al.(2014)]{2014SPIE.9147E..0LD} Dalton, G., Trager, S., 
Abrams, D.~C., et al.\ 2014, \procspie, 9147, 91470L 

\bibitem[Dawson et al.(2013)]{2013AJ....145...10D} Dawson, K.~S., Schlegel, 
D.~J., Ahn, C.~P., et al.\ 2013, \aj, 145, 10 


\bibitem[Eisenstein et al.(2011)]{2011AJ....142...72E} Eisenstein, D.~J., 
Weinberg, D.~H., Agol, E., et al.\ 2011, \aj, 142, 72 

\bibitem[]{} Everitt, B. S., Landau, S., Leese, M., \& Stah, D., 2011,  
 Cluster Analysis, 5th Edition, Wiley

\bibitem[Fern{\'a}ndez-Alvar et 
al.(2015)]{2015A&A...577A..81F} Fern{\'a}ndez-Alvar, E., Allende Prieto, C., Schlesinger, K.~J., et al.\ 2015, \aap, 577, A81 


\bibitem[]{} Garc\'{\i}a P\'erez, A. E., Allende Prieto, C., Holtzman, J. A., 2015, et al., submitted to AJ

\bibitem[Gilmore et al.(2012)]{2012Msngr.147...25G} Gilmore, G., Randich, 
S., Asplund, M., et al.\ 2012, The Messenger, 147, 25 

\bibitem[]{} Goldberg, D. E. 1989, Genetic Algorithms in Search, Optimization and Machine Learning, Addison-Wesley Longman, Boston

\bibitem[Gunn et al.(2006)]{2006AJ....131.2332G} Gunn, J.~E., Siegmund, 
W.~A., Mannery, E.~J., et al.\ 2006, \aj, 131, 2332 

\bibitem[Heiter et al.(2015)]{2015arXiv150606095H} Heiter, U., Jofr{\'e}, 
P., Gustafsson, B., et al.\ 2015, arXiv:1506.06095 


\bibitem[Hill et al.(2008)]{2008ASPC..399..115H} Hill, G.~J., Gebhardt, K., 
Komatsu, E., et al.\ 2008, Panoramic Views of Galaxy Formation and 
Evolution, 399, 115 


\bibitem[Holtzman et al.(2015)]{2015arXiv150104110H} Holtzman, J.~A., 
Shetrone, M., Johnson, J.~A., et al.\ 2015, submitted to AJ, arXiv:1501.04110 

\bibitem[Jofr{\'e} et al.(2015)]{2015arXiv150700027J} Jofr{\'e}, P., 
Heiter, U., Soubiran, C., et al.\ 2015, arXiv:1507.00027 


\bibitem[de Jong et al.(2014)]{2014SPIE.9147E..0MD} de Jong, R.~S., Barden, 
S., Bellido-Tirado, O., et al.\ 2014, \procspie, 9147, 91470M 


\bibitem[]{} Kirkpatrick, S. 1984, Journal of Statistical Physics, 34, 975

\bibitem[Kordopatis et al.(2013)]{2013AJ....146..134K} Kordopatis, G., 
Gilmore, G., Steinmetz, M., et al.\ 2013, \aj, 146, 134 

\bibitem[Lee et al.(2008)]{2008AJ....136.2022L} Lee, Y.~S., Beers, T.~C., 
Sivarani, T., et al.\ 2008, \aj, 136, 2022 

\bibitem[]{} Lee, P. M. 2012, Bayesian Statistics: An Introduction, Wiley


\bibitem[Levi et al.(2013)]{2013arXiv1308.0847L} Levi, M., Bebek, C., 
Beers, T., et al.\ 2013, arXiv:1308.0847 



\bibitem[Liu et al.(2015)]{2015RAA....15.1089L} Liu, X.-W., Zhao, G., 
\& Hou, J.-L.\ 2015, Research in Astronomy and Astrophysics, 15, 1089 


\bibitem[Ludwig et al.(2009)]{2009MmSAI..80..711L} Ludwig, H.-G., Caffau, 
E., Steffen, M., et al.\ 2009, \memsai, 80, 711 

\bibitem[Luo et al.(2015)]{2015RAA....15.1095L} Luo, A.-L., Zhao, Y.-H., 
Zhao, G., et al.\ 2015, Research in Astronomy and Astrophysics, 15, 1095 

\bibitem[Majewski et al.(2015)]{2015arXiv150905420M} Majewski, S.~R., 
Schiavon, R.~P., Frinchaboy, P.~M., et al.\ 2015, submitted to AJ, 
arXiv:1509.05420 

\bibitem[]{} Marquardt, D., 1963,  SIAM Journal on Applied Mathematics, 11, 431

\bibitem[]{} Nelder, J. A. \& Mead, R. 1965, The Computer Journal, 1965, 7, 308

\bibitem[Ness et al.(2015)]{2015ApJ...808...16N} Ness, M., Hogg, D.~W., 
Rix, H.-W., Ho, A.~Y.~Q., \& Zasowski, G.\ 2015, \apj, 808, 16 



\bibitem[Recio-Blanco et al.(2006)]{2006MNRAS.370..141R} Recio-Blanco, A., 
Bijaoui, A., \& de Laverny, P.\ 2006, \mnras, 370, 141 

\bibitem[Recio-Blanco et 
al.(2014)]{2014A&A...567A...5R} Recio-Blanco, A., de Laverny, P., Kordopatis, G., et al.\ 2014, \aap, 567, A5 


\bibitem[]{} Shewchuk, J. R. 1994, http://www.cs.cmu.edu/~quake-papers/painless-conjugate-gradient.pdf

\bibitem[Smee et al.(2013)]{2013AJ....146...32S} Smee, S.~A., Gunn, J.~E., 
Uomoto, A., et al.\ 2013, \aj, 146, 32 

\bibitem[Smiljanic et 
al.(2014)]{2014A&A...570A.122S} Smiljanic, R., Korn, A.~J., Bergemann, M., et al.\ 2014, \aap, 570, A122 


\bibitem[Steinmetz et al.(2006)]{2006AJ....132.1645S} Steinmetz, M., 
Zwitter, T., Siebert, A., et al.\ 2006, \aj, 132, 1645 


\bibitem[Trampedach et al.(2013)]{2013ApJ...769...18T} Trampedach, R., 
Asplund, M., Collet, R., Nordlund, {\AA}., 
\& Stein, R.~F.\ 2013, \apj, 769, 18 

\bibitem[Yanny et al.(2009)]{2009AJ....137.4377Y} Yanny, B., Rockosi, C., 
Newberg, H.~J., et al.\ 2009, \aj, 137, 4377 

\bibitem[York et al.(2000)]{2000AJ....120.1579Y} York, D.~G., Adelman, J., 
Anderson, J.~E., Jr., et al.\ 2000, \aj, 120, 1579 


\bibitem[Zhao et al.(2015)]{2015arXiv151008216Z} Zhao, G.-B., Wang, Y., 
Ross, A.~J., et al.\ 2015, submitted to MNRAS, arXiv:1510.08216 


\bibitem[Zucker et al.(2012)]{2012ASPC..458..421Z} Zucker, D.~B., de Silva, 
G., Freeman, K., Bland-Hawthorn, J., 
\& Hermes Team 2012, Galactic Archaeology: Near-Field Cosmology and the Formation of the Milky Way, 458, 421 


\end{thebibliography}
%



\end{document}